*Six Guidelines for Trustworthy, Ethical and Responsible Automation Design*

**Matouš Jelínek, Nadine Schlicker & Ewart de Visser**

*Abstract*

Calibrated trust in automated systems (Lee and See 2004) is critical for their safe and seamless integration into society. The common goal of the various definitions of calibrated trust[1] is to optimize outcomes in human-automation interaction. In other words, users should only rely on a system recommendation when it is actually correct and reject it when it is factually wrong. One requirement to achieve this goal is an accurate trustworthiness assessment, ensuring that the user's perception of the system's trustworthiness aligns with its actual trustworthiness, allowing users to make informed decisions about the extent to which they can rely on the system (Schlicker et al. 2022).

The field of pragmatics offers valuable insights into the communication processes essential for developing accurate trustworthiness assessments by helping to find strategies to bridging the gap between user expectations and system capabilities, emphasizing the importance of context and mutual understanding in communication.

We propose six design guidelines to help designers to optimize for accurate trustworthiness assessments, thus fostering ethical, and responsible human-automation interactions. The proposed guidelines are derived from existing literature in various fields, such as human-computer interaction, cognitive psychology, automation research, user-experience design, and ethics.

We are incorporating key principles from the field of pragmatics, specifically the cultivation of common ground (H. H. Clark 1996) and Gricean communication maxims (Grice 1975). These principles are essential for the design of automated systems because the user's perception of the system's trustworthiness is shaped by both environmental contexts, such as organizational culture or societal norms, and by situational context, including the specific circumstances or scenarios in which the interaction occurs (Hoff and Bashir 2015). Pragmatics principles provide valuable insights into these contexts, thus enhancing our understanding and design of trustworthy systems.

Our proposed guidelines provide actionable insights for designers to create automated systems that make relevant trustworthiness cues available. This would ideally foster calibrated trust and more satisfactory, productive, and safe interactions between humans and automated systems. Furthermore, the proposed heuristics might work as a tool for evaluating to what extent existing systems enable users to accurately assess a system's trustworthiness.

---

[1] We use calibrated trust here as an umbrella term that covers research on appropriate trust levels and appropriate reliance.

*Introduction*

*Significance of Trust in Human-automation Interactions*

The recent rapid development of Artificial Intelligence (AI) has made automation inevitable and transformative in a wide range of fields. AI systems are used in a range of applications, such as semi- or fully autonomous drones (Tezza and Andujar 2019), AI-based medical diagnostic systems (Panesar 2019), recommender systems shaping our media consumption, self-driving cars navigating complex traffic situations, or social robots designed to help children combat anxiety (Crossman, Kazdin, and Kitt 2018).

As intelligent agents continue to evolve, their integration across various domains is not just a possibility but a reality. On one hand, the potential benefits of automation are immense, with powerful AI systems offering different new solutions to various tasks, in some cases significantly improving efficiency and accuracy. On the other hand, the rapid deployment of AI systems introduces new challenges and risks that might hinder effective human-automation teaming or even put humans in hazardous situations (O'Neill et al. 2022).

Effective and safe human-automation interactions require users to adequately rely on the system, i.e., neither rely too much or too little in the system (Lee and See 2004). This balance is critical to avoid miscalibration that becomes apparent in phenomena like underreliance and overreliance. Underreliance, i.e., rather not relying on a system that is actually correct, arises from an underestimation of system capabilities, and may prevent users from benefiting from automation and potentially lead to complete system rejection. Overreliance, i.e., rather relying on a system that is actually incorrect, arises from an overestimation of system capabilities, resulting in phenomena like complete reliance on a system. Overreliance, influenced by the automation bias (Parasuraman and Riley 1997), increases the risk of overlooking automation errors, which in extreme cases can lead to disastrous outcomes, such as commercial aircraft crashes (Lee and See 2004), the grounding of a cruise ship or collisions caused while relying on automated systems in an autonomous vehicle (Bonnefon, Shariff, and Rahwan 2016).

One requirement to mitigate the risks of over- and underreliance is that users' expectations of the system continually align with system capabilities. Users need to correctly assess a system and its capabilities. More specifically, users need to interpret available information correctly with respect to the system's trustworthiness. We call any information element that can be used to make a trustworthiness assessment about an agent a *trustworthiness cue* (de Visser et al., 2014; Schlicker et al., 2022)

Pragmatics provides valuable insights into the design and interpretation of trustworthiness cues in human-automation interactions. The field of Pragmatics examines how meaning is constructed in context (Levinson 1983). In the context of accurate trustworthiness assessments, pragmatic theories can inform designers and practitioners how the information is understood and cognitively processed on the human side. It provides a framework for understanding the successful transmission and reception of trustworthiness cues. The theoretical insights from pragmatics are not just academic; they have practical applications for how automated systems can communicate their capabilities to users.

We propose a set of six design guidelines grounded in pragmatic theory, aimed at designers and other stakeholders involved in the interaction itself or its design. Specifically, the guidelines focus on communicating automated systems' transparency, reliability, and predictability. They range from ensuring clear communication of the system's actions, intentions, and decision-making to cautiously implementing human-like features. They address critical aspects such as system uncertainty, user control, feedback mechanisms, and error management.

Together, our approach aims to form a guiding framework for the design and evaluation of automated systems, focusing on aligning user expectations with the system's actual capabilities, thus helping to foster effective human-automation teaming.

*Pragmatics as a Framework for Trust Calibration*

Calibrated trust necessitates that users accurately interpret the trustworthiness cues provided by the system, recognizing, understanding, and evaluating the system's ability, benevolence, and integrity (Mayer, Davis, and Schoorman 1995). Schlicker and colleagues (2023), in their Trustworthiness Assessment Model (TrAM), further explore how systems must effectively communicate these cues, which is essential for bridging the gap between the perceived and actual trustworthiness of the system (E. de Visser et al. 2014). Schlicker's theory encompasses the entire process of the relevance and availability of cues, and the trustor's cue detection, interpretation, and response within their individual environment and backgrounds.

This two-sided approach aligns with Roman Jakobson's model of communication (1960), highlighting the critical role of context, code, and contact in effective communication. According to the model, the sender (addresser - the automated system in our context) communicates a message to the receiver (addressee - the user), which must operate within a context recognizable and interpretable by the receiver. The message uses a code, at least partially shared between the two parties, allowing them to encode and decode the message effectively (Jakobson 1960). For effective communication, there has to be the speaker's intention to communicate and the listener's ability to infer the intention using clues from the context, their knowledge of the world, and the conversation to figure out what the speaker means (Grice 1975).

The meaning itself is then negotiated. According to Clark (1996), effective communication discourse necessitates more than just the accurate delivery of messages. It requires a collaborative effort, a mutual belief among participants that the message has been understood well enough for current purposes. This mutual understanding, or 'common ground' is cultivated through a dynamic process where both communication partners work together to negotiate the meaning of each communication act based on adjustments, clarifications, feedback, and shared context.

*Effective Communication as the Foundation of an Accurate Trustworthiness Assessment*

Despite their current advanced capabilities, automated systems historically possess only limited contextual understanding, often failing to follow the nuanced communication patterns expected from their characters by their human communication counterparts (Clark and Fischer 2022). Still, users tend to unthinkably apply the same social rules, norms, and expectations to their interactions with automation (Reeves and Nass 1996), complicating the correct assessment of the actual degree of trustworthiness.

Designers of intelligent agents thus have to consider that the perceived trustworthiness of the system relies not only on the explicit communication of the information provided by the system but also on the implicit cues users infer from their interactions with the system. The conveyance of the system's trustworthiness is then linked to effective communication of both clear, explicit content and implications constructed by the user (inferred or suggested). This process is dynamic, and the understanding is achieved through the interplay of these cues based on their relevance to the individual's context and cognitive environment.

To help users bridge the gap between perceived and actual trustworthiness, the designers of autonomous systems should look for strategies to communicate clearly about system capabilities to avoid miscommunication. Grice's work (Grice 1975) on communication maxims offers a valuable framework for that, providing comprehensive guidance and highlighting the importance of clear, truthful, and relevant communication, which can be utilized to align user expectations with system capabilities. The maxims focus on the cooperative nature of communication, which is especially relevant when ensuring that the user accurately (or at least as accurately as possible) understands what the automated system intends to communicate:

> **Maxim of Quality:** Be truthful. Do not say what you believe to be false or that for which you lack evidence.

**Maxim of Quantity:** Be as informative as necessary. Provide as much information as needed, but not more than is required.

**Maxim of Relation:** Be relevant. Make your contribution relevant to the conversation.

**Maxim of Manner:** Be clear. Avoid ambiguity and obscurity; be brief and orderly.

We argue that by adhering to the maxims of Quality, Quantity, Relation, and Manner, designers ensure that automated systems communicate with an emphasis on mutual understanding and relevance, which is essential for effectively conveying the systems' capabilities and limitations, particularly when meaningful cooperative communication is not possible.

We have, therefore, detected six problematic areas in trust calibration throughout the interaction life-cycle previously identified in the human factors literature, and have materialized these into six design guidelines. We have informed the new guidelines by the knowledge of human cognition from the field of pragmatics, especially the Gricean communication maxims.

These guidelines resulted from a targeted theoretical synthesis, rather than a systematic methodological review, to improve trust calibration from a design perspective. Our inspiration was to create a list similar to the ten usability heuristics (Nielsen 1994) but specific to trust in automation. Accordingly, we identified ten recurring challenges through a review of the foundational literature on human-automation interaction (i.e., Lee and See 2004; Parasuraman and Riley 1997; Hoff and Bashir 2015; Hancock et al. 2011). The ten challenges were then discussed extensively and focused on removing redundancies and increasing the clarity and practicality of each corresponding guideline. The final list included the following six issues: the misalignment between user expectations and system capabilities, lack of clear explanations of system actions, difficulty in understanding system uncertainty, the need to address errors in the interaction, violation of social norms, and gaps in user training.

The selection was motivated by their prevalence in the literature, as well as their relevance to the key phases of the interaction life cycle: pre-interaction, during interaction, and post-interaction (Hoff and Bashir 2015; C. Miller et al. 2023). The resulting guidelines were developed to address these challenges and provide actionable recommendations for optimizing human-automation interactions. While we acknowledge the final list may not be exhaustive, and future research might uncover the need for additional guidelines, our goal is to highlight the most salient and recurring issues and provide a practical starting point for the designers to avoid them in the design.

Our approach highlights that for optimal human-automation interaction designers must consider the connection between language and context, recognizing that they both transform and reshape each other in a continuous loop. Designers should build systems that help establish common ground prior to the interaction, build on it during the interaction, re-establish it in case of errors or miscommunication, and evolve it post-interaction. In other words, there might be a set of expectations for automation performance before interacting with automation. Then, users interact with the automation and gain experience with the system. If errors occur, there may be ways to recover from them.

The model in Figure 1 shows how each of the guidelines maps onto this life-cycle of human-automation interaction and how it fosters optimal human-automation interaction. They can either be used to design a new automated system or to evaluate one. The next section describes each of the guidelines in more detail, with more specific design guidance, examples, and supporting evidence of the rule. The guidelines are:

Guideline 1 - Allow verification of automation operation

Guideline 2 - Explain automation actions

Guideline 3 - Convey automation uncertainty

Guideline 4 - Provide methods to recover from automation errors

Guideline 5 - Design automation to conform with social norms and etiquette

Guideline 6 - Provide automation training

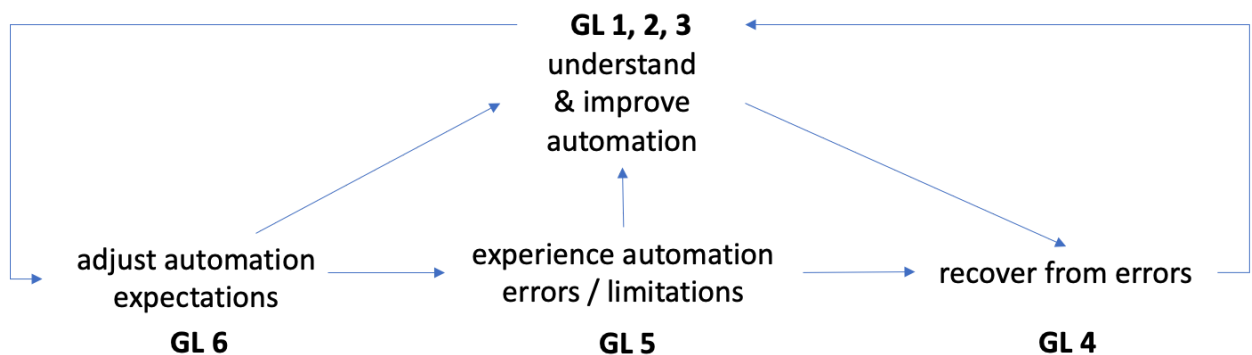

**Figure 1.** A categorization of the heuristics along the life-cycle of human-automation interaction.

*Guidelines*

*Guideline 1 - Allow verification of automation operation*

**Design Guidance:** Users must be given methods to allow for exploration, verification, and a better understanding of the system to verify underlying data that supports the automation's analysis and decision-making process. This guideline is anchored in the principles of effective communication, focusing on truthful, reliable information - driven by the Maxim of Quality, but also the Maxim of Quantity, highlighting the importance of providing enough information for verification and understanding of the systems' decision-making processes. The automated systems have to ensure that their statements are backed by evidence, allowing the users to rely on or reject the information presented in the process of negotiating the shared meaning of the cues provided.

**Supporting Evidence:** Providing operators with a means of verifying the performance of system components of an automated aid will reduce the number of errors when operators follow a recommendation of an automated aid even though this recommendation is incorrect (commission errors) (J. E. Bahner, Hüper, and Manzey 2008). The rate of how often humans tend to review the system's decisions, if they have the option (i.e. their verification rate) is related to trust and performance in supervisory automation settings (Walliser, de Visser, and Shaw 2023). Providing operators with guidance on different system modes, decision-making, and transparency of the algorithm may also help users to accurately assess system trustworthiness by constructing a shared mental model (Degani, Shafto, and Kirlik 1999). Providing aggregated and or summarized information with the option to "drill down" into data is another effective way to engender accurate trustworthiness assessments.

**Example:** Showing a video for an automated motion detection alert of a person being detected. You can check the event or the live camera to verify the alert. Similarly, in a medical scenario, explaining each step of the blood pressure measurement process can help users understand and validate the system.

**Caution:** While promoting system transparency, it is essential to balance the amount of information to avoid cognitive overload, with providing too much data or the underlying data not being meaningful or informative to the actual context. For example, by providing continuous automation performance feedback but not showing every single automation decision caused more appropriate reliance when an automated aid exhibited superior performance (Dzindolet et al., 2003). This is because obvious automation errors can cause decreased trust (P. Madhavan and Wiegmann 2007; Poornima Madhavan, Wiegmann, and Lacson 2006; De Visser et al. 2016).

Fischer et al. (2018) highlighted that transparency needs to be context-sensitive, providing enough information to build trust without overwhelming the user or exposing unnecessary technical details that might reduce trust if they reveal system limitations excessively.

*Guideline 2 - Explain automation actions*

**Design Guidance:** The system should aim at continuously improving users' understanding, to help set appropriate expectations about what the system is doing, how it's doing it, and how well it's doing it, and hence to reach informed decisions. Grice's Maxims help to understand what makes a good explanation. First, explanations should be true and not contain anything for which evidence is lacking or is knowingly false. Second, explanations should be appropriate in terms of quantity, neither too little nor too much information should be provided. Third, explanations should be tailored to the needs of the receiver. Fourth, an explanation should be genuine and not spurious. That means it should be clear and well organized. Fifth, the explanation should be conveyed politely.

**Supporting Evidence:** Explanations are only helpful if they are detected and utilized correctly by the trustor (Gajos and Mamykina 2022; Buçinca, Malaya, and Gajos 2021). That means that trustors need to engage with the explanation, process the provided information, and hence reach an informed, justified decision (Ferrario and Loi 2022). As a result, explanations should help users form a correct mental representation of the system to make appropriate predictions about the system's performance (Lee and See 2004; Endsley 2023; Rouse and Morris 1986). This should enable users to differentiate the right from the wrong system recommendations. Note that "the user" might come in very different forms, with different levels of expertise and acting in various contexts with a large variability of explanation requirements (Langer et al. 2021). Research showed that interaction quality and empathy also increased acceptance of virtual agents (Pelau, Dabija, and Ene 2021).

**Example:** Using audio and visual cues, a robo-taxi informs its passengers it will change lanes or take a turn before it does so.

**Caution:** Much work highlighted the importance of creating transparent and explainable automated systems (Chen et al. 2018; Arrieta et al. 2019), however, not every explanation does result in an accurate trustworthiness assessment and appropriate reliance (T. Miller 2023; Langer et al. 2021). Some studies show negative effects of explanations, with people relying more on incorrect AI advice when an explanation is present (Jakubik et al. 2022; Rieger et al. 2023). Designers should consider who is the receiver of the explanation and the kind of information that different user groups may need to improve their understanding of the system. Further, they should evaluate whether the given explanation met the goal of increasing user understanding, adhering to the Gricean Maxim of Quality.

*Guideline 3 - Convey automation uncertainty*

**Design Guidance:** While striving to design systems that are as error-free as possible, the systems should transparently and truthfully communicate about the automation confidence, adhering to Grice's Maxim of Quality. This includes honestly highlighting the limitations of the system, which allows users to better assess the capabilities of the system and anticipate when intervention is necessary.

**Supporting Evidence:** Adhering to the Maxim of Manner is critical to ensure the communication is clear and orderly. Communication of uncertainty should be put in an appropriate context, i.e. to clearly explain to users what the confidence measures mean and how it should be interpreted. Conveying automation uncertainty directly is an effective method to optimize human-automation interaction by providing machine confidence indicators (Beller, Heesen, and Vollrath 2013; Helldin et al. 2013). Visualizing uncertainty might help users correctly interpret uncertainty. For instance, decision information icons (DICONS) are a way to show dimensions of uncertainty such as completeness and conflict (E. de Visser et al. 2014). Degrading icons can be used to convey situational uncertainty (Finger and Bisantz 2002). Uncertainty in paths that automated vehicles take could be displayed using ensemble displays that combine many option paths in one visual display (Liu et al. 2015). Likelihood alarms are effective at providing graded indications about the reliability of signal detection measures using sensitivity (Sorkin, Kantowitz, and Kantowitz 1988; Yang et al. 2017).

**Example:**

- In a heavy rain, autonomous cars provide the following information: "Sensor visibility reduced to 70%. Be prepared to take manual control if needed."

**Caution:**

System confidence is not a clear-cut case, and the confidence provided is not always a good indicator for its performance. Bach (2023) showed that visual certainty cues might increase efficiency in decision-making but lower the chance of humans recognizing an AI error if the system is quite confident. Therefore, system designers should be aware of the consequences of a "go" sign and probably adjust the threshold to increase accuracy. Examples of confidence miscalibrations of AI systems were for instance found when introducing noise in adversarial attacks (Heaven 2019). People also struggle with interpreting uncertainty (Spiegelhalter, Pearson, and Short 2011) and confidence scores (Du, Huang, and Yang 2020). This is partially caused by the imperfection in the way the various scales are used.

*Guideline 4 - Provide methods to recover from automation errors*

**Design Guidance:** Trust repair strategies should be in place in a system to provide resilience against inevitable automation errors. The role of communication in trust repair is crucial. Designers should enable the automated systems to send clear signals of their intentions and corrective actions to their users. Such communication should be clear, relevant, and timely, reflecting the context and expectations, and adhering to Grice's Maxims of Quality and Manner.

**Supporting Evidence:** Types of automation errors such as false alarms and misses both have detrimental and unique effects on trust and the joint performance of human-automation teams (McBride, Rogers, and Fisk 2014; Rice and McCarley 2011; Sanchez et al. 2014; Wickens and Dixon 2007). False alarms can build distrust and consequently lead to disuse of the system (Parasuraman and Riley 1997). Different strategies can be used to mitigate these errors including trust repair and dampening approaches (E. J. de Visser et al. 2020). Trust repair approaches attempt to restore lost trust due to false alarms or other causes. These methods may include different types of apologies, explanations, or assurances from the automation (E. J. de Visser et al. 2020). Trust can be further repaired by re-designing the system, or having a trusted party convey their confidence in the system (trust propagation).

**Example:** A robotic assistant, after mistakenly interrupting a user during a session, would say "Oh, sorry, I think I interrupted you. My microphone isn't working well today. I'm trying to do better. Could you repeat what you were saying before I interrupted you?" (Axelsson, Spitale, and Gunes 2024), addressing the error and communicating the immediate corrective action.

**Caution:** Be aware that not all errors can be fixed by simple apologies. Apology attempts could backfire if the user perceives them as insincere or repetitive without evident improvement (Zechmeister et al. 2004), such attempts may further erode the level of trust rather than restore it.

*Guideline 5 - Design automation to conform with social norms and etiquette*

**Design Guidance:** Automation should conform to the social norms of the situation and context in which it is deployed. Adhering to human etiquette and social norms leverages the Gricean Maxims of Manner and Relation. Designers should ensure that the system's communication is not only contextually appropriate, but also socially acceptable, predictable, clear, and avoids ambiguity. Adding human-like features to the system may help counter existing biases and rectify false expectations. However, these features must be implemented carefully to avoid creating unrealistic expectations about system capabilities.

**Supporting Evidence:** Automation that adheres to social norms, such as politeness or etiquette increases trust (Brown and Levinson 1987; Hayes and Miller 2011; Parasuraman and Miller 2004). Early research on automation demonstrated that automation deployed in human settings can violate human social norms. A computer can be considered "rude" if it interrupts people or does not appear to listen to what people say. Conversely, polite, non-interruptive systems are trusted more (Parasuraman and Miller 2004). Moreover, this politeness effect increased trust in unreliable automation to the point that it was considered more trustworthy than even rude reliable automation!

Benefits of incorporating human-like features were found regarding e.g. a system's likeability, its perceived intelligence, the user's willingness to disclose (Lucas et al. 2014) trust, acceptance, and empathy, and the joint-performance of humans and systems within a task (Roesler, Manzey, and Onnasch 2021), and also regarding improved trustworthiness and acceptability of the system (Verberne, Ham, and Midden 2012). Adding human-like features to an interface can even increase trust resilience (De Visser et al. 2016; Pak et al. 2012).

Even if the designer decides not to introduce additional human-like features users may tend to anthropomorphize a system. As such amplifying this human tendency to anthropomorphize by adding human-like features (e.g. appearance, social behavior, empathy, voice (Pelau, Dabija, and Ene 2021) or by simply adding a human-like framing (e.g., giving it a human name or a human description (Onnasch and Roesler 2019; Kopp, Baumgartner, and Kinkel 2023) was found to initiate human-like interaction schemes.

**Example:** A virtual assistant in an elderly-care facility scenario engages with a resident by politely asking, "How are you feeling today?" to start the interaction. Immediately after, it clearly states its limitations, transparently communicating its boundaries, saying, "Remember, I can assist by calling a nurse or providing some general information, but please note I am an AI system, not a healthcare professional."

**Caution:** The incorporation of human-like features can set high expectations (Reeves and Nass 1996) concerning the system's understanding, empathy, or adaptability. If these expectations do not match the capabilities of the system, trust can quickly decline. Designers must ensure that these expectations align with the system's actual capabilities to avoid confusion and inappropriate trust or consider reducing anthropomorphism (Culley and Madhavan 2013). Designers should also be aware of the uncanny valley effect (Mori, MacDorman, and Kageki 2012), suggesting that there might be a spot between clearly artificial and clearly human that could induce perceptions of eeriness. Further, different cultural backgrounds lead to varying expectations and interpretations of social norms and etiquette in automation (Chien et al. 2016). It is crucial to design automation that conforms to diverse cultural norms and etiquettes to foster an effective human-automation interaction. Also, the effect of anthropomorphization on joint task performance is not fully understood and depends on the context and the task (Roesler, Manzey, and Onnasch 2021).

*Guideline 6 - Provide automation training*

**Design Guidance:** Training should be an integral part of the human-automation lifecycle, designed to set clear and appropriate expectations about the system's capabilities and limitations. It is essential to minimize automation surprises by establishing a common ground (H. H. Clark 1996), that includes awareness of the system's characteristics, its behavior in different situations, and potential failures, thus avoiding future communication problems. Training helps to make the specifics of the system's "language and culture" clear to potential users with different languages and cultures (Cummings 2019).

**Supporting Evidence:** Effective training transitions users from active decision-makers to competent overseers of automation. That means that trainees have enough knowledge and ability to oversee a task, are aware of potential loss of situational awareness, and also that their intentions are aligned with that of the task (e.g. that intervening may produce additional efforts for them and the team, but still that they are empowered to check properly if they doubt a system's decision) (Sterz et al. 2024). Previous work has demonstrated that exposing operators to automation failures during training reduced complacency (J. Bahner, Elepfandt, and Manzey 2008). Dzindolet et al. (2003) showed that providing a rationale for why the aid may fail increased trust and ultimate reliance on the aid. It is therefore important to know the nature of errors, why they occur, and what they mean in the context of the automated aid. Goal setting and role clarification successfully improved trust and performance in a human-machine teaming context (Walliser et al. 2019). Providing training with decision aids has resulted in increased understanding of the aid and operators learning new ways to interact with it (Cohen et al. 1997). There have also been calls to provide certifications for learning how to drive with automated systems (Cummings, 2019).

**Example**: After each autonomous car update, the user has to familiarize themselves with the new features and confirm that they understand them. The car might show a message such as: "Welcome to your updated driving system. Please note that the update may change how the system reacts in certain situations. It's important to review the following changes and how the system can fail (examples). Always be prepared to take over driving."

**Caution:** The AI Act (European Parliament 2024) suggests that humans interacting with an AI system need to be enabled to effectively oversee it—for example, by deciding when (not) to use it and being aware of the biases they may fall for. Training is only successful if users are able to interact with the system afterwards and are able to rationally justify their (mis)trust in the system in different circumstances. Operators also need to be properly informed and possibly retrained if system updates are introduced.

*Discussion*

*Contributions*

This paper applies pragmatics theory to long-standing human factors challenges to provide new insights into optimal human-automation interaction. It addresses recent calls to embrace the responsible use of AI as part of the human-centered design process, as opposed to abstract calls for creating "a more ethical AI" (Lyons et al. 2023; Pflanzer et al. 2023). Applying pragmatics theories to the challenges of human-automation interaction provides new insights. Specifically, the sender-receiver model, which describes how messages are sent and received, can be reconciled with the TrAM model, which describes the differences between perceived trustworthiness (receiver) and actual trustworthiness (sender). We further converted these insights into six practical guidelines that can be used to improve human-automation and human-autonomy interaction.

*The Need for Trustworthiness Heuristics*

Recent calls for trustworthy and responsible AI (Jobin, Ienca, and Vayena 2019) have become vital as concerns about AI has risen to the top of policy agendas around the world. The recent EU law governing AI outlines several ways AI should ensure human oversight and control. Yet, these laws are too general to serve as practical design guidelines. This legislation functions more like a list of requirements. A translation needs to occur to convert these requirements into an implementable AI design. This initiative is consistent with recent calls for "trust engineering" in AI and human-autonomy teams (Dorton and Stanley 2024; Ezer et al. 2019).

There may be some serious design trade-offs that present themselves when implementing trustworthy AI. For example, more monitoring of AI could present a higher burden for human operators responsible for the AI which defeats the purpose of AI in alleviating human workload. This is known as the "automation paradox" or the "double-edge sword of automation" (Parasuraman and Wickens 2008). Designers need to carefully implement how the human supervises and checks the AI without increasing the burden for human workload. Another example is the lumberjack effect (Onnasch et al., 2014) which poses that with increases in autonomy, failures have more catastrophic consequences. All this poses considerable design challenges, and specific guidance must be given to translate these high-level principles to practical, implementable design guidance. Our paper presents a first step towards this goal.

*How Previous Lists Are Not Adequate*

Previous lists have described and characterized the need for improved human-machine interaction. For example, there are 10 heuristics to assess the usability and design of a system (Nielsen 1995), 10 challenges for making automation a better team player (Klien et al. 2004), a list that describes what machines are better and what humans are better at (de Winter and Dodou 2014), a list of design considerations in the life-cycle of human-automation interaction (C. A. Miller 2021), and there are ten commandments of software that describe how to create truly agile software development ("DIB Guide: Detecting Agile BS" 2018).

We believe our list of guidelines still contributes uniquely to the literature for several reasons. First, our guidelines are specifically designed to enhance human-automation interaction based on pragmatics principles of human communication, such as those outlined by H. H. Clark (1996) and Grice (1975). These principles provide a solid basis for human interaction, emphasizing common ground and effective communication, which are critical for human-agent interaction. Second, these guidelines are based on research from the last four decades on trust in automation. These guidelines are based on actual findings that show what specific challenges arise when interacting with AI and the difficulty of addressing these. By leveraging pragmatic principles, our guidelines ensure that the communication between humans and automated systems is clear, truthful, and contextually appropriate, fostering appropriate reliance.

Third, our work translates these research findings into a concrete guideline that can help designers improve their human-automation designs. This translation is necessary to aggregate results across research findings into a coherent, logical, and practical guideline. Fourth, our list provides a framework of where in the human-automation interaction process these guidelines may apply. Lastly, this list cautions where the implementation or design can fail if implemented clumsily and provides guidance on how to avoid poor implementation.

We believe that, while our work may be a promising start to articulate a set of important guidelines, this list is not exhaustive. More rules may be appropriate. Some of the guidelines are rooted in entire fields of study including explainable AI, transparent automation, and politeness research. This may require more detailed sub-guidelines to guide design at a granular level and as new research within this field emerges. In addition, new research may also reveal trade-offs between guidelines and even the original maxims. For example, creating a polite system may promote trust for users, but it can reduce efficiency by adding more words and ambiguity, which contradicts the Maxims of Quantity and Manner.

*Implications and Future Research Directions*

An important implication of this work for pragmatics theory may be to improve mutual understanding between humans and automation, which in turn can improve human-automation interaction. We term this idea "mutually adaptive trust calibration" a process which occurs when both the automated agent and machine agents continuously update their beliefs, attitudes, and behaviors to better adapt to one another (Castelfranchi and Falcone 2010; E. J. de Visser et al. 2023; Chen et al. 2018). Mutually adaptive trust calibration could be considered a form of AI alignment. In this work, we primarily focused on helping to improve the design of automated systems for human designers. Conversely, the automated system can adapt to the human. Systems that exhibit this kind of behavior are known as adaptive trust calibration systems (E. J. de Visser et al. 2020; Okamura and Yamada 2020) and they adapt to the user to better understand human attitudes, beliefs, and behaviors. With the recent emergence and development of foundational models, such machine understanding of humans may improve dramatically. Other recent efforts have focused on the idea of AI alignment, a form of adaptive automation, where the AI systems are programmed to align with human values or key decision-maker attributes with the goal of creating a system that exhibits the desired behavior that is consistent with human values.

Our work has shown that it can be valuable to look into the field of Pragmatics, which may provide additional value to enhance communication between humans and automation. Besides improving humans' ability to understand automation and automation's ability to understand humans, there is a need to develop more dynamic and fluent real-time communication. Human-human communication is flexible, real-time and fluent coordination. Real communication is naturally messy and requires quick adjustments, adaptation to errors, and recognition of mutual understanding. Current systems using natural-language approaches are static, turn-based, and do not exhibit memory beyond a few speech turns. Therefore, a more flexible and dynamic approach is needed that can account for errors, can adjust on the fly and can calibrate in real-time. The ultimate version of success with these approaches is to establish bi-directional and adaptive mutual trustworthiness assessments.

*Conclusion*

Informed by the fields of human-automation interaction and pragmatics, we present six design guidelines to improve the trustworthiness of automated systems. The implementation of these guidelines may improve human-automation interaction and aid in the design of trustworthy, ethical, and responsible automation.

*Future Research*

Future research should focus on validating the guidelines through empirical studies in real-world scenarios. For example, experiments could test the effectiveness of the guidelines in fostering calibrated trust across domains. Additionally, practitioners' feedback can be vital in refining the guidelines and uncovering practical challenges and opportunities for further adaptations and refinement of the guidelines.

As for now, the proposed guidelines are intended to serve as both a checklist and a source of inspiration for designers, highlighting what to prioritize, how to proceed, and what risks to anticipate during the design process. In addition, the next step can be the development of more detailed checklists or step-by-step guidelines that can further guide practitioners in applying these guidelines effectively.